# A Study on the Line of Sight to Galaxies Detected at Gamma-ray Energies

Amy Furniss,[1] Josepf N. Amador,[2,3] Olivier Hervet,[1] Ollie Jackson,[1] and David A. Williams[1]

[1]*Santa Cruz Institute for Particle Physics and Department of Physics*
*University of California, Santa Cruz*
[2]*Department of Physics and Astronomy*
*San José State University, San José, California*
[3]*Department of Physics and Astronomy*
*California State Polytechnic University, Pomona*

## ABSTRACT

The large-scale Universal structure comprises strands of dark matter and galaxies with large under-dense volumes known as voids. We measure the fraction of the line of sight that intersects voids for active galactic nuclei (AGN) detected by *Fermi* Large Area Telescope (LAT) and quasars from the Sloan Digital Sky Survey (SDSS). This "voidiness" fraction is a rudimentary proxy for the density along the line of sight to the galaxies. The voidiness of SDSS-observed quasars (QSOs) is distinctly different from randomly distributed source populations, with a median p-value of $4.6 \times 10^{-5}$ and $\ll 1 \times 10^{-7}$, when compared with 500 simulated populations with randomly simulated locations but matching redshifts in the $0.1 \leq z < 0.4$ and $0.4 \leq z < 0.7$ intervals, respectively. A similar comparison of the voidiness for LAT-detected AGN shows median p-values greater than 0.05 in each redshift interval. When comparing the SDSS QSO population to the LAT-detected AGN, we mitigate potential bias from a relationship between redshift and voidiness by comparing the LAT-detected AGN to a "redshift-matched" set of SDSS QSOs. The LAT-detected AGN between a redshift of 0.4 and 0.7 show higher voidiness compared to the redshift-matched SDSS QSO populations, with a median p-value of $2.3 \times 10^{-5}$, (a $4.1\sigma$ deviation). No deviation is found when comparing the same populations between redshifts of 0.1 and 0.4 ($p > 0.05$). We do not study possible causes of this voidiness difference. It might relate to propagation effects from lower magnetic or radiative background fields within voids or to an environment more favorable for gamma-ray production for AGN near voids.

## 1. INTRODUCTION

Very high energy (VHE; $E \geq 100$ GeV) gamma rays can interact with cosmological fields when traveling extragalactic distances. The gamma rays interact with the extragalactic background light (EBL) through pair production (Nikishov 1962; Gould & Schréder 1967), which allows the density of the photon field to be indirectly probed through observation of gamma-ray emitting AGN (e.g. Biteau & Williams (2015); H. E. S. S. Collaboration et al. (2017); Fermi-LAT Collaboration et al. (2018); Biasuzzi et al. (2019); Abeysekara et al. (2019); Acciari et al. (2019)). The pairs resulting from this interaction are deflected by the intergalactic magnetic field (IGMF; Aharonian et al. (1994); Neronov & Semikoz (2009); Alves Batista & Saveliev (2021)). The electron-positron pairs may initiate plasma instabilities in the thermal background (Broderick et al. 2012; Schlickeiser et al. 2012; Miniati & Elyiv 2013; Perry & Lyubarsky 2021; Alawashra & Pohl 2024) and can inverse-Compton upscatter cosmic microwave background (CMB) photons to gamma-ray energies. The combination of all three interactions provides indirect means to probe the magnitude of the IGMF through the observation of extended GeV emission in the direction of gamma-ray emitting AGN (e.g. Archambault et al. (2017); H. E. S. S. Collaboration et al. (2014); Acciari et al. (2023)) or in the time delay from short-term transient events such as gamma-ray bursts (GRBs; e.g. Vovk et al. (2024); Abdalla et al. (2022)) or flaring gamma-ray blazars, e.g. (Dermer et al. 2011; Acciari et al. 2023).

Furniss et al. (2015) investigated the possible effect of EBL inhomogeneity within voids. A simulated column-shaped void between the Milky Way and a galaxy at $z = 0.6$ was found to have a $\sim 2\%$ lower internal EBL density compared to outside of the column. A similar result was found by Abdalla & Böttcher (2017).

**Table 1.** Summary of the SDSS QSO and 4LAC catalogs for which voidiness was calculated. 4LAC AGN without redshift information were removed from the study. All sources in the SDSS QSO catalog have redshift values, and duplicate sources with equivalent location were removed. The number of sources within nearby ($0.1 \leq z < 0.4$) and distant ($0.4 \leq z < 0.7$) redshift intervals overlapping the Sutter et al. (2012) and Sutter et al. (2014) void catalog coverage are provided.

| Population | Reference | Catalog Non-Duplicated Total | $z < 0.7$ | $0.1 \leq z < 0.4$ | $0.4 \leq z < 0.7$ |
|---|---|---|---|---|---|
| 4LAC DR3 | Ajello et al. (2022) | 3,472 | 328 | 160 | 143 |
| SDSS QSOs | Lyke et al. (2020) | 797,606 | 19,796 | 3,326 | 16,425 |

Furniss et al. (2015) characterized the lines of sight to VHE-detected and VHE-candidate AGN from the first *Fermi*-LAT hard source catalog (1FHL; Ackermann et al. (2013)), resulting in a marginal indication that VHE-like and Fermi hard sources were detected along lines of sight which had a larger fraction intersecting voids than randomly distributed points. This result was based on the limited statistics of the 19 VHE-like and 28 hard Fermi sources which fell within the void survey volume from Sutter et al. (2012), which extends to a redshift of $z = 0.36$.

In this work, we measure the fraction of the line of sight to SDSS-detected and gamma-ray-detected AGN intersecting voids measured in the redshift range from 0.1 to 0.7 (Sutter et al. 2012, 2014).

## 2. THE OPTICAL AND GAMMA-RAY CATALOGS

The galaxies included in this study are taken from two catalogs (summarized in Table 1). Specifically, we characterize the lines of sight to quasars (denoted as QSOs) within the catalog derived from SDSS DR16 (Lyke et al. 2020), removing 22,359 redundant catalog entries through identification of QSOs with duplicate right ascension (RA) and declination (Dec) values. We also characterize the line of sight to AGN reported in the *Fermi* LAT 4th Catalog of Active Galactic Nuclei (Ajello et al. (2020), hereafter referred to as 4LAC), specifically in the DR 3 catalog which is based on 12 years of LAT data (Ajello et al. 2022). The breakdown of source class for the 4LAC AGN included in the study is 15 flat-spectrum radio quasars (FSRQ), 130 BL Lacertae (BL Lac) objects, 12 blzar candidates of unknown type (BCU), 1 narrow-line seyfert 1 (NLSY1) and 2 radio galaxies in the redshift range of $0.1 \leq z < 0.4$; and 27 FSRQs, 94 BL Lacs, 17 BCU, 1 compact steep spectrum (CSS) quasar, 1 SSRQ (a class of non-blazar AGN) and 3 NLSY1 in the redshift range of $0.4 \leq z < 0.7$. The line of sight calculations were completed for catalog sources which fall within the volume of the void survey. Voidiness was calculated for sources within the SDSS DR 7 and 9 footprint at $0.1 \leq z < 0.7$. We utilize updated redshift information for 4LAC sources included in Goldoni et al. (2021).

The galaxy populations were filtered to fall within the void catalog footprint using the `alphashape` Python package which creates a polygon surrounding the outermost points of interest (void locations, specifically). The resulting boundary was inspected visually to confirm included galaxies were within the void catalog volumes reported by Sutter et al. (2012, 2014).

## 3. LINE OF SIGHT CHARACTERIZATION

The Sutter et al. (2012) and Sutter et al. (2014) void catalogs span the redshift ranges of $0 < z < 0.44$ and $0.36 < z < 0.7$, respectively. The voids are measured with the void-identification methods of modified and extended versions of ZOBOV (Neyrinck 2008) and VIDE (Sutter et al. 2015), applied to the SDSS DR7 and DR9 data, respectively (Sutter et al. 2012, 2014). This study utilizes the catalog RA, Dec, redshift and radii of 5143 total voids.

The fraction of the line of sight which passes through a void, referred to as "voidiness" in Furniss et al. (2015), is denoted as $V$ and defined as

$$V = \frac{D_{\text{void}}}{D_{\text{LoS}}} \quad ,$$

where $D_{\text{LoS}}$ is the total comoving distance to the galaxy and $D_{\text{void}}$ is the sum of the line of sight segments which intersect voids. The intersecting distance $D_{\text{void}}$ is the magnitude of the mathematical union of the segments within voids so that intersections of overlapping voids are not double-counted.

## 4. TREATMENT OF POTENTIAL REDSHIFT-VOIDINESS BIAS

The potential for a relationship between galaxy redshift and voidiness is investigated separately for each population. The median voidiness and the one and two standard deviations from each median were calculated in redshift bins of width 0.1 for $0 < z < 0.7$, with each bin containing sources across the full 0 to 1 range of possible voidiness. In the 0.1 to 0.7 redshift range, the median voidiness per bin fell between 0.27 and 0.40 for the 4LAC AGN, and 0.30 and 0.37 for the SDSS QSOs, with the two standard deviation full width in each bin measuring between 0.3 and 0.4 for 4LAC and between 0.4 and 0.6 for the SDSS QSOs. Notably, for the limited number of AGN at $z < 0.1$ in both the 4LAC and SDSS catalogs (25 and 45, respectively), the median voidiness was ∼15% lower, with values of 0.12 and 0.18, respectively, which is still consistent with, but lower than, the measured standard deviations of the higher $z$ bins. A systematically lower voidiness for sources at $z < 0.1$ could result from the void-identification algorithm which requires an intervening filament of galaxies to be present for convergence. This results in nearly half of the nearest galaxies having voidiness close to zero. To avoid this possible bias, we do not use galaxies with $z < 0.1$ in the subsequent analysis.

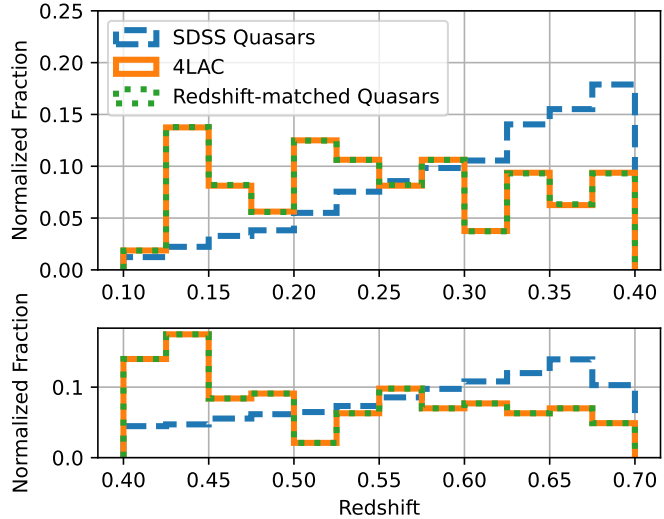

**Figure 1.** The redshift distributions of the SDSS QSOs (blue dashed line) and 4LAC AGN (orange line) within the footprint of the Sutter et al. (2012, 2014) void catalogs. A redshift-matched SDSS QSO population (dotted green) is also shown.

To further mitigate any low-level redshift-voidiness bias, we split the study into two redshift intervals, referred to as nearby $(0.1 \leq z < 0.4)$ and distant $(0.4 \leq z < 0.7)$. Smaller redshift intervals were not investigated, limited by the number of 4LAC sources (160/143) in the nearby/distant intervals, respectively.

The redshift distributions for the galaxies included in this study are in Figure 1. We produce 500 "redshift-matched" populations of SDSS QSOs to match the redshift distribution of the 4LAC AGN by randomly pulling QSOs from the full SDSS QSO distribution to create a population with a redshift distribution that matches that of the 4LAC. Due to the very different redshift distributions of the 4LAC and SDSS QSOs, we only compare the voidiness of the 4LAC and "redshift-matched" QSO populations.

We additionally produce 500 simulated populations for each of the 4LAC and SDSS QSO populations separately, referred to as the randomized populations hereafter. These randomized populations contain the same number and redshifts of galaxies as their respective catalogs, but with the RA and Dec randomly produced and assigned so that the simulated galaxies fall within the celestial coordinates of the survey region from which the void-catalogs were built.

## 5. RESULTS

The comparison of two populations is completed with a two-sample Kolmogorov-Smirnov (KS) test using the Python module `scipy.stats.ks_2samp`, which returns a KS-statistic and p-value (as described in, e.g., Kendall & Stuart (1979)). The returned p-value corresponds to the probability that the two tested samples are drawn from the same underlying continuous distribution. We quote the median p-value and KS-statistic when comparing observed sources to each of the 500 randomized simulations individually. All KS test p-values are crosschecked with the ones resulting from a k-sample Anderson-Darling (AD) test using the Python module `scipy.stats.anderson_ksamp`. The AD test (Anderson & Darling 1952) has been estimated as superior to the KS test in multiple studies, especially when samples differ in their distribution tails (e.g. Stephens 1974). As the AD Python module returns p-values capped at $p \in [1 \times 10^{-3}, 0.25]$, most of the KS test p-values are checked in regard to the lower/upper limits given by the AD test.

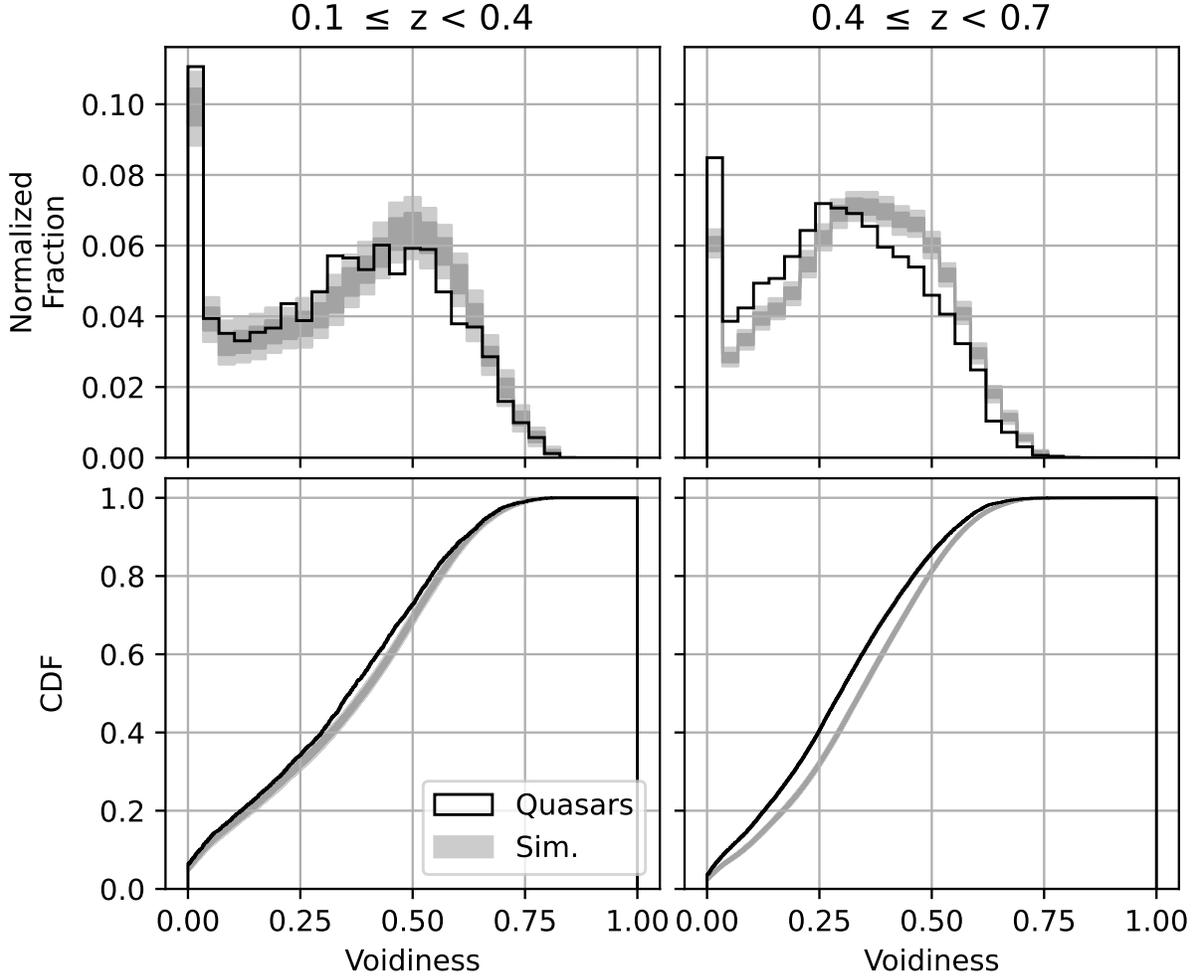

**Figure 2.** Top panels: The voidiness distributions of the observed SDSS QSOs (black line), shown along with the contours for one and two standard deviations from the median of the randomized populations (grey shaded regions), calculated per displayed voidiness bin in the nearby (left) and distant (right) redshift intervals, respectively. Bottom panels: The CDFs for the observed SDSS QSOs (black line) and contours one and two sigma from the median of the randomized populations in the nearby (left) and distant (right) redshift intervals. The data, which include the source name, location, redshift and calculated voidness, are available in the online journal as a data behind the figure (dbF) file.

The results for the SDSS QSOs and randomized populations are shown with the voidiness distribution and corresponding cumulative distribution function (CDF) in Figure 2. The contours show the one and two standard deviations from the median of the randomized populations calculated within each displayed voidiness bin. The medians of the KS-statistic and corresponding p-values are 0.056 and $p = 4.6 \times 10^{-5}$ ($p_{AD} < 1 \times 10^{-3}$) for nearby QSOs and 0.095 and $p \ll 1 \times 10^{-7}$ ($p_{AD} < 1 \times 10^{-3}$) for distant QSOs.

Figure 3 shows the voidiness distribution and CDF for the 4LAC AGN and 4LAC-specific randomized populations. In comparing the 4LAC AGN to the randomized populations, the medians of the KS-statistics were 0.11 and 0.13 in the nearby and distant intervals, respectively. The medians of the p-value, confirmed with AD test, were found to be greater than 0.05 for both the nearby and distant intervals.

The 4LAC AGN were also compared to redshift-matched populations of the SDSS QSOs. The median of the KS-statistics resulting from these comparisons were 0.076 and 0.2 in the nearby and distant redshift intervals. For the nearby interval, the median p-value (KS and AD tests) was greater than 0.05, while for the distant interval $p = 2.3 \times 10^{-5}$ ($p_{AD} < 1 \times 10^{-3}$). This median p-value for the distant redshift interval indicates that the population of 4LAC AGN has lines of sight with substantially different voidiness compared to SDSS QSOs with similar redshifts, at a level of $4.1\sigma$. Considering the three largest subsets of the 4LAC AGN population separately, BL Lacs (94 objects), FSRQs (27 objects), and BCUs (17 objects), we find the significances of the difference in voidiness distribution with respect to the SDSS QSO sample to be $3.8\sigma$, $0.3\sigma$, and $2.7\sigma$, respectively. The BL Lac sources are responsible for the bulk of the effect, especially if the majority of the BCU objects, which are unidentified so far, are in fact BL Lacs.

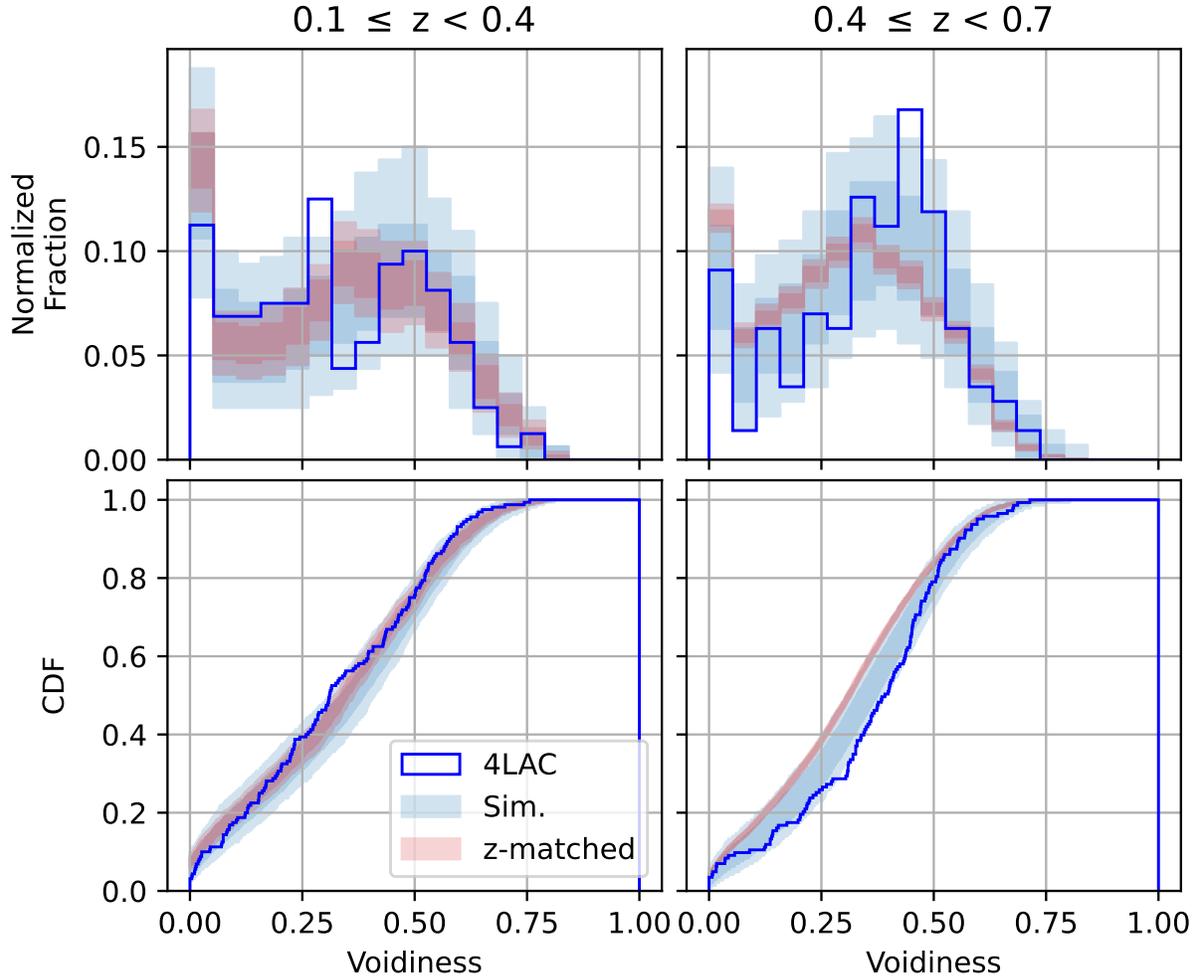

**Figure 3.** Top panels: The voidiness distributions of the 4LAC AGN (blue line), shown along with the contours one and two standard deviations from the median of the randomized populations (blue shaded regions), calculated in each voidiness bin in the nearby (left) and distant (right) redshift intervals, respectively. Bottom panels: The CDF for the observed 4LAC voidiness (blue line) and the one and two sigma contours from the median of the randomized populations, calculated in each displayed voidiness bin. In all panels the redshift-matched SDSS QSO populations' one and two standard deviation contours are shown in red shading. The data, which include the source name, location, redshift and calculated voidiness, are available in the online journal as a data behind the figure (dbF) file.

## 6. DISCUSSION

We present evidence that SDSS-detected QSOs are not distributed in the Universe in the same way as randomly placed locations in the survey volume. This is consistent with the observation that galaxies are found along filamentary strings of dark matter (Bond et al. 1996).

The observation that distant ($0.4 \leq z < 0.7$) gamma-ray-detected AGN reported in the 4LAC show a different distribution of voidiness compared to SDSS QSOs has a statistical significance of $4.1\sigma$. The deviation is specifically skewed toward gamma-ray-detected AGN showing higher voidiness than SDSS QSOs, which suggests the former lie along lines of sight which have lower average density. There is no indication that a similar deviation occurs for more nearby ($0.1 \leq z < 0.4$) 4LAC AGN.

We checked for catalog biases affecting the KS-test results. In addition to the potential bias arising from a redshift-void correlation (removed using SDSS "redshift-matched" samples), we checked for a potential correlation between the voidiness of 4LAC sources and the sensitivity map of the *Fermi*-LAT from the 4FGL-DR3 (Abdollahi et al. 2022). We found a modest negative correlation (*Fermi*-LAT was slightly more sensitive in the directions to sources with lower voidiness) at $1.5\sigma$ confidence level, excluding any *Fermi*-LAT detection bias toward sources with higher voidiness.

We are not aware of any prediction of the effect we have observed, which is therefore unexpected. Nevertheless, there are plausible mechanisms which might produce the result. A lower EBL density in voids could result in fewer gamma-rays being absorbed during their propagation to Earth. This effect has been estimated to be small (Furniss et al. 2015; Abdalla & Böttcher 2017), but may deserve further investigation. Another possible explanation of this observation is that distant 4LAC sources are being detected with a measurable number of secondary gamma-ray photons resulting from primary VHE photons interacting with the EBL. This interaction results in pairs which can be deflected by the IGMF and upscatter CMB to the LAT energy range. These interactions make it possible for the observed gamma-ray flux of these sources to be dependent on the magnitude of the intervening magnetic field. A lower magnitude of magnetic field within voids compared to outside of voids could lead to a higher level of gamma-ray flux observed along lines of sight with higher voidiness, but such effects are expected to be rather small, e.g. Finke et al. (2015). Finally, the environment of a void may be advantageous for gamma-ray production, favoring gamma-ray emission from AGN that are in or near voids. Detailed, quantitative calculations, beyond the scope of this letter, will be needed to establish whether these mechanisms or any other can account for the observed effect.

We would like to thank Paul Sutter, Jonathan Biteau, and Manuel Meyer for valuable discussions about this work. We also thank the referees, whose suggestions significantly improved this paper. We gratefully acknowledge support from the Cal-Bridge Program for summer support of Josepf Amador, as well as support from the U.S. National Science Foundation awards PHY-2110974, PHY-2310002, and PHY-2411760.